\documentclass[traditabstract]{aa} 
\usepackage{graphicx,natbib} 
\usepackage{txfonts} 
\bibpunct{(}{)}{;}{a}{}{,} 
 
\def\kms{km~s$^{-1}$}
\def\jyb{Jy~beam$^{-1}$}

\begin{document} 
 
\title{Polarization of thermal molecular lines in the envelope of IK~Tau} 
 
\author{Vlemmings, W. H. T.\inst{1} 
\and Ramstedt, S.\inst{2} \and Rao, R.\inst{3} \and Maercker, M.\inst{2,4}}

\institute{Department of Earth and Space Sciences, Chalmers University of Technology, Onsala Space Observatory, SE-439 92 Onsala, Sweden
\email{wouter.vlemmings@chalmers.se} 
\and Argelander Institute f\"ur Astronomie, Universit\"at Bonn, 
Auf dem H\"ugel 71, 53121 Bonn, Germany\\ 
\and Submillimeter Array, Academia Sinica Institute of Astronomy and
Astrophysics, 645 N. Aohoku Place, Hilo, HI 96720, USA \\
\and European Southern Observatory, Karl Schwarzschild Str. 2, Garching bei M\"unchen, Germany}
\date{Received: 27 January 2012 , Accepted: 12 March 2012} 
\authorrunning{Vlemmings et al.} 
\titlerunning{Molecular line polarization around IK~Tau} 
 
\abstract {Molecular line polarization is a unique source of
  information about the magnetic fields and anisotropies in the
  circumstellar envelopes of evolved stars. Here we present the first
  detection of thermal CO($J$\,=\,2$\rightarrow$1) and
  SiO($J$\,=\,5$\rightarrow$4, $\nu$\,=\,0) polarization, in the
  envelope of the asymptotic giant branch star IK~Tau. The observed
  polarization direction does not match predictions for circumstellar
  envelope polarization induced only by an anisotropic radiation
  field. Assuming that the polarization is purely due to the
  Goldreich-Kylafis effect, the linear polarization direction is
  defined by the magnetic field as even the small Zeeman splitting of
  CO and SiO dominates the molecular collisional and spontaneous
  emission rates. The polarization was mapped using the Submillimeter
  Array (SMA) and is predominantly north-south. There is close
  agreement between the CO and SiO observations, even though the CO
  polarization arises in the circumstellar envelope at $\sim800$~AU
  and the SiO polarization at $\lesssim250$~AU. If the polarization
  indeed traces the magnetic field, we can thus conclude that it
  maintains a large-scale structure throughout the circumstellar
  envelope. We propose that the magnetic field, oriented either
  east-west or north-south is responsible for the east-west elongation
  of the CO distribution and asymmetries in the dust envelope. In the
  future, the Atacama Large Millimeter/submillimeter Array will be
  able to map the magnetic field using CO polarization for a large
  number of evolved stars.}
 
\keywords{polarization, magnetic field, Stars: AGB} 
 
\maketitle 
 
\section{Introduction} 

Molecular line polarization observations can provide invaluable insights
into the magnetic field and/or anisotropies of the circumstellar
environment of asymptotic giant branch (AGB) stars. If even a
relatively weak magnetic field is present, the molecular gas in the
envelope of the AGB star will show linear polarization when the
magnetic sublevels of the rotational states are exposed to 
anisotropic emission (the Goldreich-Kylafis effect, Goldreich \&
Kylafis 1981, 1982). Alternatively, \citet{morris85} predict polarized
line emission arising from molecules in an envelope that have a
preferred rotation axis because of radial infrared emission from the central
star. In this case, deviations from radial symmetry in the
polarization angles could be due to a non-spherically symmetric envelope and/or
to a (non-radial) magnetic field. While polarization observations
of molecular lines, in particular of CO, in star-forming regions are
relatively common \citep[e.g.][]{Cortes05, Beuther2010}, the only
previously published significant detection of molecular line
polarization around an AGB star is that of the
CS($J$\,=\,2$\rightarrow$1) transition around IRC+10216
\citep{Glenn1997}. 

Most information about the magnetic fields in the envelopes
of AGB stars is currently derived from maser observations
\citep[e.g.][]{Vlemmings2005, Herpin2006}. The maser observations
reveal strong magnetic fields throughout the entire envelope
\citep[e.g.][]{Vlemmings2011}. As maser observations typically probe
only a limited number of lines-of-sight through the stellar envelope,
the polarization of thermal molecular lines can more easily reveal the
circumstellar magnetic field morphology. A strong magnetic field is,
in addition to both the binary and disk interactions, one of the possible mechanisms
for shaping the typically spherical AGB winds into an aspherical
planetary nebula \citep[e.g.][]{balick02, frank07}.

Here we present Submillimeter Array (SMA) polarization observations of
the molecular lines in the circumstellar envelope of IK~Tau, which is
a well-studied M-type AGB star. It has a period of 500 days
\citep{Kuk1971}, and a fairly high mass-loss rate of
1$\times$10$^{-5}$\,M$_{\sun}$\,yr$^{-1}$ \citep{Ramstedt2008}. The
star has a rich circumstellar chemistry, as indicated by the detection
of a large number ($>$20) of different molecular species \citep[see
e.g., ][for the latest results]{Kim2010,Decin2010}. The circumstellar
gas distribution has been mapped in CO($J$\,=\,1$\rightarrow$0),
CO($J$\,=\,2$\rightarrow$1) \citep{Castro2010}, and
CO($J$\,=\,3$\rightarrow$2) \citep{Kim2010}
emission. \citet{Castro2010} found a western elongation in its
innermost region (of $\sim1000$~AU), and on larger scales a
flattened circumstellar envelope that is elongated in the east-west plane
over $\sim10.000$~AU. Elongated, elliptical circumstellar structures
are also found on scales from a few to a few tens of AU in maps of the
SiO \citep[e.g., ][]{Cotton2010} and H$_{2}$O \citep{Bains2003} maser
emission. The SiO maser observations indicate that there is a magnetic field close
to the central star of $1.9$-$6.0$~Gauss \citep{Herpin2006}.

\section{Observations and data reduction} 
 
The observations of IK~Tau were done with the SMA on October 8 2010 in
the compact configuration with the lower- and upper-side band (LSB and
USB) covering the frequency ranges of $216.8$-$220.7$~GHz and
$228.9$-$232.8$~GHz respectively. This frequency setting was
specifically selected to cover the CO($J$\,=\,2$\rightarrow$1)
transition at $230.538$~GHz, but also covered transitions of SiO, SiS,
and SO. We list the detected molecular transitions in
Table~\ref{lines}.  The correlator setup provides a spectral
resolution of $\sim0.8$~MHz, which corresponds to $\sim1.0$~\kms~ at
$230$~GHz. At $230$~GHz, the beam size is
$\sim3.0\times2.6$~arcseconds with a position angle of
$81.4^\circ$. The observations lasted for a total track of nine hours
and included observations of the phase calibrator J0423-013, bandpass
and polarization calibrator 3C454.3, and the primary flux calibrators
Callisto and Neptune. Based on the flux measured for J0423-013
($\sim2.2$~\jyb), we estimate that the total flux uncertainty is
$\lesssim 10\%$. The right ascension ($\alpha$) and declination
($\delta$) of the phase center for the observations of IK~Tau were
taken to be $\alpha_{\rm J2000}=03^h53^m28^s.84$ and $\delta_{\rm
  J2000}=+11^\circ24'22$\arcsec$.56$. The data were calibrated
initially in the MIRIAD software package using the specific
polarization calibration technique \citep{Marrone2008}.  Using a
standard SMA polarization observing schedule, the polarization
calibrator 3C454.3 was observed regularly, covering $>120^\circ$ in
parallactic angle. The gain calibrator J0423-013 is linearly
  polarized, but does not show any circular polarization. The derived
  amplitude gain corrections thus do not affect the amplitude
  calibration of the right- and left-circular polarizations produced
  by the SMA observation.  The antenna leakage solutions were found
to be up to $2\%$ in the USB, near the CO($J$\,=\,2$\rightarrow$1)
line and up to $5\%$ in the LSB, near the SiO($J$\,=\,5$\rightarrow$4,
$\nu$\,=\,0) line.  The linear polarization, $P_l$, and electric
vector polarization angle (EVPA) determined on 3C454.3 (LSB:
$P_l=0.71\pm0.12\%$, EVPA$=-47\pm6^\circ$; USB: $P_l=0.80\pm0.09$,
EVPA$=-52\pm2^\circ$) indicate that after calibration the remaining
leakage is $<0.5\%$. After calibration of the flux, phases, bandpass
and polarization, we exported the data into the Common Astronomy
Software Applications (CASA) package and
performed self-calibration on the strongest spectral channel of the
SiO line in the LSB. Although the SiO emission is only marginally
resolved, we used the initial total intensity image as a model. This
assumes that the SiO emission has no significant circular polarization,
which was verified to be the case. The SiO self-calibration solutions
were compared with self-calibration solutions derived from the
strongest spectral channel of the CO line (also following an initial
image as a starting model) in the USB and were found to be
identical. We then applied the SiO self-calibration solutions to both
LSB and USB as the SiO solutions were obtained using data with higher
signal-to-noise ratios. Maps of the total intensity Stokes I and the linear
polarization intensities Stokes Q and U were created using 'briggs'
weighting. To optimize the polarization detection, spectral channels
were averaged over an interval of $5$~\kms~ for the CO, SiS, and SO lines and $2.5$~\kms~ for
the SiO lines. As a results, the channel root-mean-square (rms) noise
ranges from $\sim35$~mJy~beam$^{-1}$ for the CO line to
$\sim65$~mJy~beam$^{-1}$ for the SiO line. We note that the noise in the
LSB is $\sim30\%$ higher than that in the USB. We also imaged the
  CO line without self-calibration in order to test the effect of
  self-calibration on the polarization. The EVPAs and linear
  polarization structure were found to be the same as when
  self-calibration was applied. Only the significance of the polarization
  detections decreased by $\sim15\%$.

\begin{table}
\caption{Properties of observed lines}
\begin{tabular}{l|c|c|c|c}
      \hline
      \hline
%      Line        & Frequency & Peak Flux & Integrated Flux & Max. Pol
 %    & Extent \\
      Line        & $\nu_0$ & $F_p$ & $F_i$ & $P_l^{\rm max}$ \\
             & [GHz] & [\jyb] & [\jyb & [$\%$]\\
             &  &   & \kms] & \\
   \hline
     SiO($J$\,=\,5$\rightarrow$4, $\nu$\,=\,0) & 217.105 & 15.2 & 379 & $8.8$ \\
     SiS($J$\,=\,12$\rightarrow$11)   & 217.818 & 3.7 & 93 & $<5.4$ \\
      SO($J$\,=\,6$\rightarrow$5) & 219.949 & 2.0 & 58 & $<6.2$ \\
      CO($J$\,=\,2$\rightarrow$1)  &  230.538  & 7.5 & 204 & $13.1$ \\
   \hline
\end{tabular}
\label{lines}
\end{table}

%\section{Results} 
%\subsection{Molecular line polarization} 
\section{Results} 
\label{linepol}

\begin{figure*}
\centering
\resizebox{0.7\hsize}{!}{\includegraphics{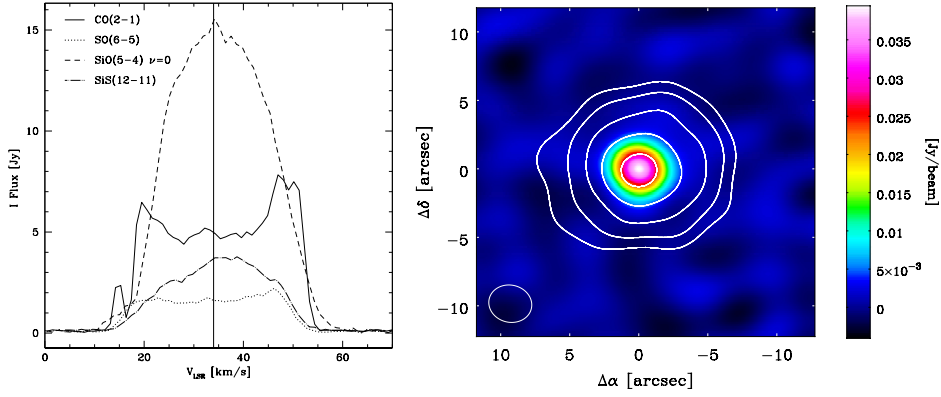}}
\caption{(left) Spectra of the observed molecular lines in the
  envelope of IK~Tau. The emission is summed over all SMA
  baselines. The vertical line indicates the $V_{\rm LSR}=34$~\kms~
stellar velocity. (right) The stellar continuum and dust emission (color) and integrated
CO(2-1) line emission (contours) of IK~Tau. The contours are drawn at
$10, 20, 40, 80,$~and~$160$~\jyb~\kms.}\label{specdust}
\end{figure*}

\begin{figure*}
\centering
\resizebox{0.75\hsize}{!}{\includegraphics{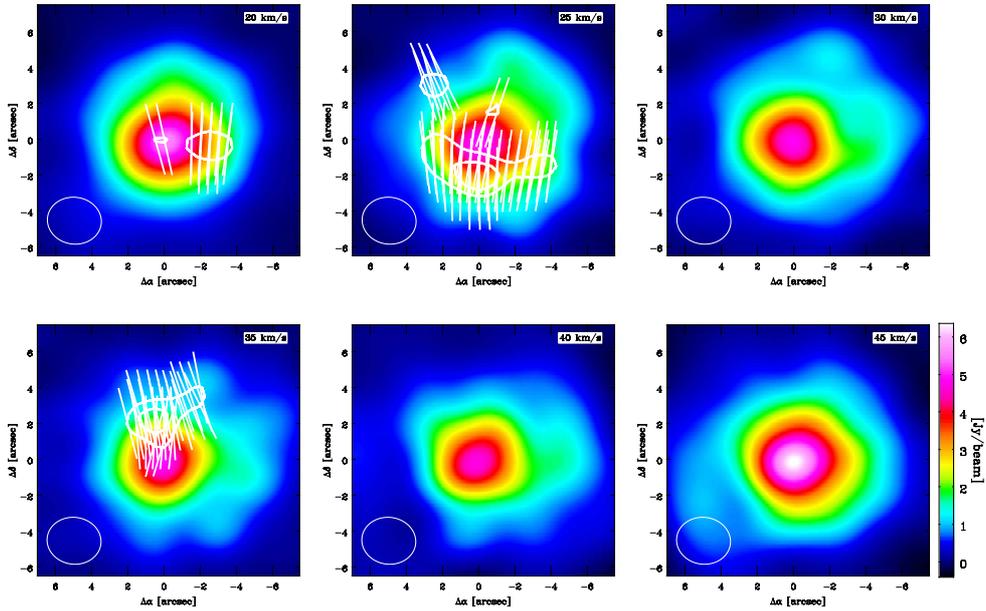}}
\caption{The polarization of the CO(2-1) line at 230.538~GHz in the
  circumstellar envelope of IK~Tau. Channels are averaged across
  intervals of $5$~\kms~
  width and are labeled according to the velocity at the lower end
  of the spectral bin. The color indicates the CO emission and the
  contours are the linearly polarized intensity. Contours are drawn
  at a significance of $3$ and $4\sigma$ (debiased, see text). The
  line segments indicate the electric vector polarization angle
  (EVPA). The beam size is indicated in the lower left corner of each panel.}\label{copol}
\end{figure*}

\begin{figure*}
\centering
\resizebox{0.9\hsize}{!}{\includegraphics{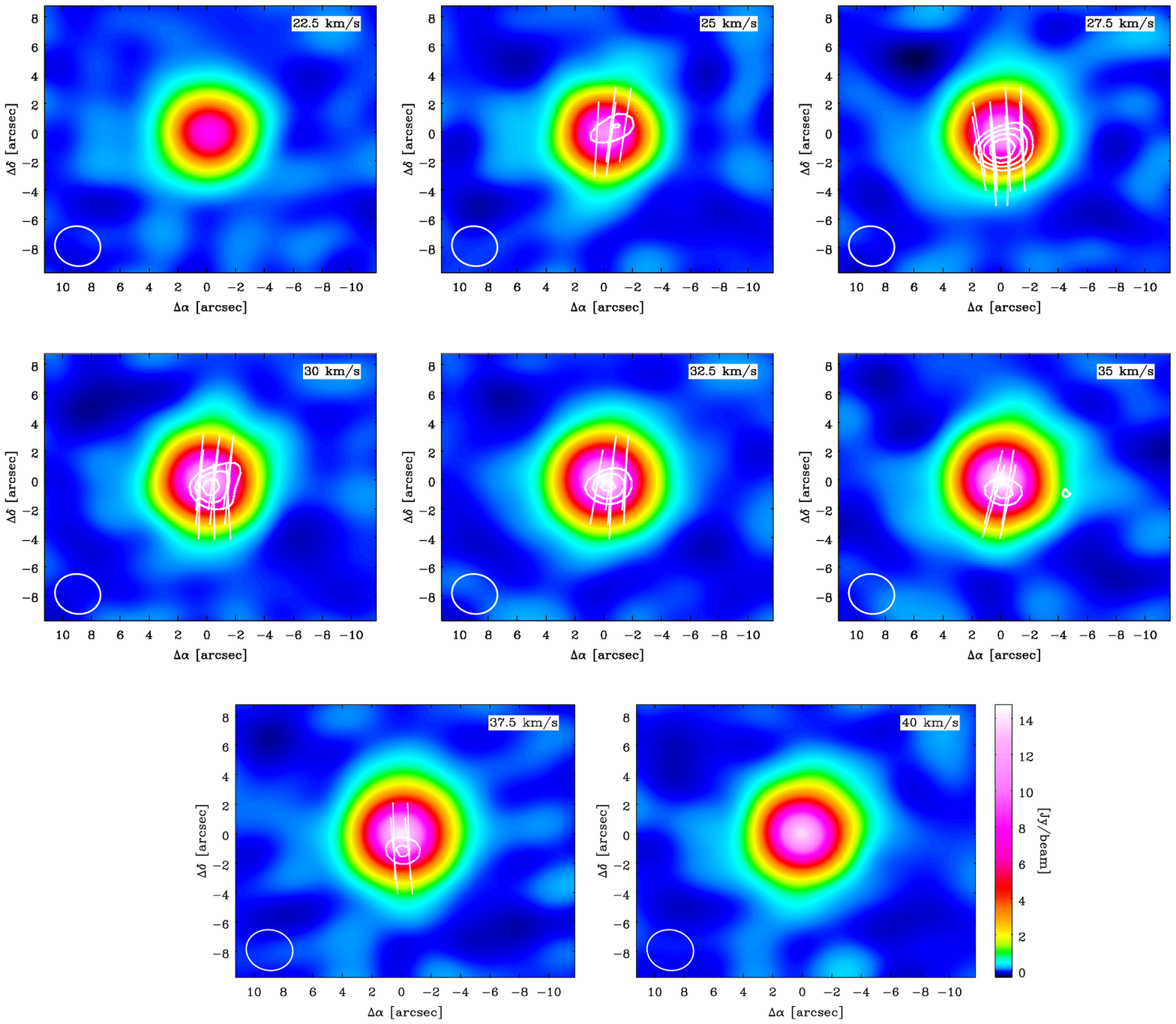}}
\caption{As Fig.\ref{copol} for the SiO(5-4) $\nu=0$ line at
217.105~GHz. In this case, channels are averaged over intervals of $2.5$~\kms~ and
contours are drawn at debiased polarized intensity levels of $3, 4,
5,$ and $6\sigma$.}\label{siopol}
\end{figure*}

Emission was detected from the CO($J$\,=\,2$\rightarrow$1),
SiO($J$\,=\,5$\rightarrow$4, $\nu$\,=\,0), SO($J$\,=\,6$\rightarrow$5),
and SiS($J$\,=\,12$\rightarrow$11)  transitions, hereafter indicated by CO(2-1),
SiO(5-4), SO(6-5), and SiS(12-11). The spectra of these transitions is
shown in Fig.~\ref{specdust}~(left). Significant polarization was
detected for the CO(2-1) and SiO(5-4) lines shown in
Fig.~\ref{copol} and online Fig.~\ref{siopol}. To determine the significance of
the weak linear polarization signal, we carefully removed the positive
bias, which was introduced when the Stokes Q and U measurements were combined to produce the linearly
polarized emission ($P_l=\sqrt{Q^2+U^2}$), following
\citet{wardle74}. As the errors in $P_l$ are not Gaussian-distributed, the
significance is defined by determining the probability intervals. The
contours in Fig.~\ref{copol} and online Fig.~\ref{siopol} thus correspond to the
$3\sigma=99.73\%$ and higher probability intervals. The peak of the
linear polarization corresponds to $4.9\sigma$ and $6.8\sigma$
detections for the CO(2-1) and SiO(5-4) lines, respectively. The maximum
polarization fractions are listed in Table.\ref{lines}. The EVPAs for both CO(2-1) and SiO(5-4) are
predominantly oriented north-south. The polarization angle of CO(2-1)
ranges from approximately $-10^\circ$ in the southern part of the
envelope in the channel around $V_{\rm LSR}=25$~\kms~ to $\sim10^\circ$
in the northern part in the channels with $V_{\rm LSR}=25$ and
$35$~\kms. A somewhat smaller spread ($-7$ to $4^\circ$) is seen for the
SiO(5-4) line, although for this line the polarized emission is confined close to the
peak of the Stokes I emission. The measurement uncertainties for the
EVPA ranges from $6$-$10^\circ$ for CO(2-1) and $4$-$10^\circ$ for
SiO(5-4). This indicates that the difference in EVPA between the
northern and southern part of the CO envelope is potentially real.

%\subsection[]{Continuum emission}
%\label{cont}
%
%We also imaged the unresolved continuum emission from IK~Tau in both
%the LSB and USB separately. The USB map is presented together with the
%integrated CO(2-1) intensity contour map in
%Fig.~\ref{specdust}~(right). The integrated continuum flux corresponds
%to $44.0\pm1.5$~mJy and $43.2\pm1.4$~mJy for the LSB and USB
%respectively and agree within the estimated $10\%$ flux uncertainties
%between the two bands. We estimate the contribution to this flux from
%a stellar black body by assuming a distance $D=265$~pc, an effective
%temperature $T_{\rm eff}=2200$~K, and a stellar radius
%$R_*=1.5\times10^{13}$~cm \citep{Decin2010b}. We then find a stellar
%continuum contribution of only $3.4$~mJy and $3.8$~mJy for the LSB and
%USB.  However, stellar pulsations introduce uncertainties in
%$T_{\rm eff}$ and $R_*$ that can be as large as $50\%$. Still, a
%significant fraction of the continuum flux appears to be due to the
%circumstellar dust. Based on these number, no dust continuum
%polarization was detected down to a $3\sigma$ level of $\sim3\%$.

\section{Discussion} 
\label{discussion}

\subsection{Polarized molecular line emission}
\label{polem}

The models of polarized molecular line-emission prescribe that the
polarization can be caused by a preferred radiation direction
\citep{morris85} or an interaction with a magnetic field
(Goldreich \& Kylafis 1981, 1982). In the first case, the stellar
  radiation will lead to polarization predominantly perpendicular to the
  radial direction. When the magnetic field is non-radial, and strong
  enough to determine the molecular alignment axis, competition
  between the field and radiation direction will change the
  polarization characteristics. \citet{morris85} state that for any
  field stronger than a few $\mu$G, the magnetic field in the envelope
  determines the molecular alignment axis. For the Goldreich-Kylafis
effect, the linear polarization is oriented along the magnetic field
as long as the Zeeman splitting dominates the collisional and
spontaneous emission rates.

The importance of the magnetic field in determining the polarization
characteristics of molecular line emission was estimated in
\citet{Kylafis1983}. We used their Eqs. 1, 2, and 3, with the
parameters adjusted to describe the CO(2-1) transition, and assumed a
magnetic field strength $\sim1~\mu$G in the CO(2-1) region, a magnetic
moment $\mu=-0.27\mu_N$ (with $\mu_N$ the nuclear magneton), and a
reduced dipole matrix element $d=0.11$~debye. It is then clear that a
hydrogen number density $\lesssim3\times10^6$~cm$^{-3}$ is sufficient
for the magnetic field to determine the direction of the polarized
CO(2-1) emission. As the SiO(5-4) emission originates from closer to
the star, the stronger magnetic field will also result in a dominant
Zeeman splitting. Owing to the field strength measured around
IK~Tau and other AGB stars ($>1$~mG within several hundred AU,
\citet{Vlemmings2005}), the linear polarization thus likely traces the
magnetic field morphology when assuming the polarization originates from
the Goldreich-Kylafis effect. As indicated in \citet{Kylafis1983} the
  CO EVPAs, arising from the Goldreich-Kylafis effect, are determined
  by the magnetic field direction on the plane of the sky. This is
  shown irrespective of the direction of the velocity gradient
  in the envelope, except that the relative optical depth caused by
  the velocity gradient determines whether the EVPAs are parallel or
  perpendicular to the magnetic field direction.

The observed linear polarization fraction is, however, larger than
predicted from the general Goldreich-Kylafis effect. \citet{Cortes05}
predict $\lesssim3\%$ for CO(2-1) when the optical depth is low
($\tau\sim1$). They also show that higher fractional polarization can
be achieved when a strong anisotropic radiation source is
present. However, in that case, the polarization direction is
determined by the anisotropic radiation field that, in the case of a
circumstellar envelope, would cause tangential polarization. As the
linear polarization in the envelope of IK~Tau is neither tangential
  nor purely radial as predicted by \citet{Deguchi1983} from
  multi-level calculations involving infra-red excitation, it is 
unlikely that the anisotropic stellar radiation field contributes
significantly. The high polarization fraction is potentially caused by
other anistotropies in the circumstellar enviroment. More detailed
modeling will be required to determine the origin of the large
fractional polarization.

\subsection{The magnetic field of IK~Tau}

For both the CO(2-1) and SiO(5-4) lines, we observe polarization that is
slightly offset from the peak of the total intensity emission. For
CO(2-1), the largest polarization fraction is found in the south part of the
blue-shifted side of the envelope and in the north part of the
red-shifted side. The observed structure in the polarized
  emission is likely an effect of similar structure in the optical depth,
  as the largest fractional polarization is found when the optical
  depth $\tau\sim1$.

As the circumstellar magnetic field strength has been shown to be
  sufficient to determine the molecular alignment axis, the
  polarization vectors are either parallel or perpendicular to the
  magnetic field direction in the framework outlined by
    \citet{Kylafis1983}. In that case, the overall field geometry is
  predominantly either east-west or north-south. As only relatively
few independent polarization vectors are measured, a more accurate
magnetic field reconstruction will require more sensitive
observations. We are able to compare the EVPA direction derived
from the CO(2-1) and SiO(5-4) EVPAs with the previous measurements
made of the SiO maser polarization by \citet{Herpin2006}. In that
paper, the 86~GHz SiO
%($J$\,=\,2$\rightarrow$1, $\nu$\,=\,1) 
maser linear polarization, averaged over several individual maser
features with the 30-m IRAM telescope, was found to have an EVPA of
$170$-$180^\circ$ for the masers blue-shifted with respect to the
star. The EVPA then rotates to $140$-$160^\circ$ for the red-shifted
masers. Thus, the CO(2-1) and SiO(5-4) EVPAs are, for the blue-shifted
emission, remarkably consistent with the averaged 86~GHz SiO
masers. However, for the 43~GHz SiO masers, \citet{Cotton2010}
  found, from high-angular resolution interferometric observations a
  much more complex and variable polarization structure.  The
CO(2-1) polarized emission is detected about $3\arcsec$
($\sim800$~AU) from the star and that of SiO(5-4) originates within
  $\sim1\arcsec$($\sim250$~AU), which could be an indication that, as
in the case of the supergiant VX~Sgr \citep{Vlemmings2011}, the
magnetic field has a large-scale component that is preserved
throughout the envelope. As the field is then mainly oriented
  either east-west or north-south, it might be directly related
to the slight east-west extent seen in the circumstellar CO envelope
\citep[][and Fig.~\ref{specdust}(right)]{Castro2010} and the east-west
asymmetry in the dust distribution
\citep{Weiner2006}. Magneto-hydrodynamical simulations indicate, for
example, that a dipole magnetic field in a circumstellar envelope can
result in equatorially enhanced density profiles
\citep{Matt2000}. However, we note that the elliptical SiO
maser structure is oriented more in a northeast-southwest direction
\citep{Cotton2010}.

\subsection{Prospects for ALMA}
Observations such as the ones presented here will undergo a
revolution when the Atacama Large Millimeter/submillimeter Array
(ALMA) becomes fully operational. With 50 antennas and in typical
weather conditions, ALMA will reach the same sensitivity in the
linear polarization as reached with the SMA (assuming similar angular
resolution and spectral binning) in less than two
seconds. In particular for IK~Tau, it will be able to improve the
observations to a $10\sigma$ linear polarization detection with higher
spatial resolution ($1" \times 1"$) and velocity resolution ($2$~\kms)
in less than 20~min on-source observing time. While the SMA
observations filter out a large fraction of the CO(2-1) flux, ALMA
with the compact array and total power antennas will recover most of
the emission and reach a similar polarization detection level even for
the evolved stars with weaker CO emission. Thus, on the basis of the SMA
observations presented here, with ALMA it will be possible to map out
the CO(2-1) and other molecular-line linear polarization in great
detail for a large number of (post-)AGB stars and planetary nebulae in
a relatively short time. As a consequence, ALMA polarization
observations will provide unique information on the magnetic field
morphology around evolved stars.

\section{Conclusions} 
\label{conclusions}

We have presented the first observations of polarized
CO($J$\,=\,2$\rightarrow$1) and SiO($J$\,=\,5$\rightarrow$4,
$\nu$\,=\,0) emission in the circumstellar envelope of an evolved
star.  
%Polarization of CO($J$\,=\,3$\rightarrow$2) in the envelope of
%the AGB star IRC+10216 has also recently been detected (Girart et al.,
%in prep.).  
As the polarization is neither radial nor tangential, the linear
polarization does not match typical predictions for linear
  polarization induced solely by an anisotropic radiation field or
  optical depth effects. It thus potentially traces the magnetic
field even when influenced by anisotropic emission from the central
AGB star IK~Tau. In that case, the magnetic field is oriented
either east-west or north-south and a good correspondence is found
between the direction derived from the CO polarization at $\sim800$~AU
from the central star and that from the thermal SiO at
$\sim250$~AU. The slight east-west elongation of the CO and the
previously observed dust asymmetry could be related to the large-scale
magnetic field morphology. Further observations and theoretical
  work are needed to properly understand the influence of the
  magnetic field and the envelope and radiation anisotropies on the
  circumstellar linear polarization. It may then be possible to
  fully describe the circumstellar magnetic field structure. In
particular, ALMA will be able to uniquely map out polarization for a
significant sample of AGB and post-AGB objects.

%______________________________________________________________ 
 
\begin{acknowledgements} 
  The authors thank the anonymous referee for comments that
  significantly improved the paper. This research was supported by the
  Deutsche Forschungsgemeinschaft (DFG; through the Emmy Noether
  Research grant VL 61/3-1).
\end{acknowledgements} 
 
%______________________________________________________________ 

\Online

\end{document}